\documentclass[aps,twocolumn,showpacs]{revtex4}

\newcommand{\be}{\begin{equation}}
\newcommand{\ee}{\end{equation}}
\newcommand{\bea}{\begin{eqnarray}}
\newcommand{\eea}{\end{eqnarray}}

\usepackage{graphicx}

\begin{document}

\title{Curvature-Induced Defect Unbinding in Toroidal Geometries}

\author{Mark Bowick,$^{1}$ David R. Nelson,$^{2}$ and Alex Travesset$^{3}$}

\affiliation{$^1$ Physics Department, Syracuse University,
Syracuse NY 13244-1130, USA \\}

\affiliation{$^2$ Lyman Laboratory of Physics, Harvard University,
Cambridge MA 02138, USA\\}

\affiliation{$^3$ Physics Department, Iowa State and Ames Natl.
Lab., Ames IA 50011-3160, USA\\}

\begin{abstract}
Toroidal templates such as vesicles with hexatic bond
orientational order are discussed. The total energy including
disclination charges is explicitly computed for hexatic order
embedded in a toroidal geometry. Related results apply for tilt or
nematic order on the torus in the one Frank constant
approximation. Although there is no topological necessity for
defects in the ground state, we find that excess disclination
defects are nevertheless energetically favored for fat torii or
moderate vesicle sizes. Some experimental consequences are
discussed.

\end{abstract}

\maketitle

\section{Introduction}

Amphiphilic molecules in water or oil solutions have been
intensely investigated over the last decade in a variety of
settings. Amphiphiles in aqueous solution, under appropriate
conditions, have been experimentally observed to form vesicles
with the topology of the sphere, torus \cite{MuBe:91} and even
higher genus surfaces \cite{MiBe:95}. Other experimental studies
have focused on the phases of amphiphilic films as a function of
temperature. It is by now well established that the high
temperature fluid phase goes into a smectic-C $L_{\beta'}$ phase
through an intermediate rippled $P_{\beta'}$ phase.  Within the
$L_{\beta'}$ phase itself there are several other phases
characterized by the degree of tilt and hexatic bond order
\cite{SSSC:88,SSSPC:90}. This beautiful experimental work may
provide insight into biological problems such as membrane fusion
\cite{Sieg:99}, where it has been argued \cite{MaAl:02,KoKo:02}
that molecular tilt plays an important role.

A remarkable understanding of the shapes of fluid amphiphilic
systems has been provided by physical methods based on the
Helfrich Hamiltonian \cite{Helf:73} and its variations
\cite{JSL:93,Seif:97}. In \cite{LuPr:92}, the problem of
fluctuating smectic-C membranes, previously investigated in
\cite{SeNe:89} for planar films, was addressed and predictions for
the shape as a function of the elastic constants were presented.
It was found that toroidal vesicles were favored for some
parameters. It was assumed, however, that free disclinations are
energetically unfavorable and may therefore be ignored unless, as
in the case of the sphere, topological constraints require them.

The main result of this paper is that disclinations can be
energetically favored over a wide range of parameters, even when
not required by topological constraints. We treat vesicles that
are topologically torii (closed surfaces of genus one). The
Gauss-Bonnet theorem for torii requires a vanishing total
disclination charge. We assume hexatic order in the tangent plane
of the torus, arising from an anisotropic liquid phase of
molecules with zero shear modulus. If $(x_1,x_2)$ are coordinates
on the torus, specified in three dimensions by a function
$\vec{R}(x_1,x_2)$, local hexatic order can be described by a bond
angle field $\theta(x_1,x_2)$ (up to rotations by
$2\pi/6=60^{\circ}$) defined relative to the local tangent vectors
$\vec{e_1} \propto
\partial_1\vec{R}$ and $\vec{e_2} \propto
\partial_2\vec{R}$. On the torus, this pair can be chosen to be
nonsingular and orthonormal everywhere;
$\vec{e_i}\cdot\vec{e_j}=\delta_{ij}$. As discussed, e.g., by
David \cite{David:89}, the usual hexatic energy on such a curved
surface can be written \be \label{hexen} E= \frac{1}{2}K_A\int
d^2x\sqrt{g}\,(\partial_i\theta - \Omega_i)(\partial_j\theta -
\Omega_j)g^{ij} \ , \ee where $K_A$ is the hexatic stiffness
constant, $g^{ij}(x_1,x_2)$ is the inverse of the metric tensor
\be \label{metten} g_{ij} = \frac{\partial\vec{R}}{\partial x_i}
\cdot \frac{\partial\vec{R}}{\partial x_j} \ee and $g=\det
g_{ij}$.

The vector-potential-like function $\Omega_j(x_1,x_2)$ in
Eq.(\ref{hexen}) describes the geometric frustration which arises
when vector or tensor order parameters are parallel transported on
curved surfaces. This ``spin connection" can be computed from
covariant derivatives acting on $\vec{e_1}$ and $\vec{e_2}$; the
``curl" of $\Omega$ (when appropriately defined on a curved
surface) is proportional to the local Gaussian curvature
\cite{David:89}.

Disclinations can be inserted into the free energy (\ref{hexen})
just as in flat space. If $N$ defects with charges $q_j=\pm1$ are
present on the torus, we first define a defect density $s$ as a
function of $x\equiv(x_1,x_2)$, namely \be\label{defden} s(x) =
\frac{2\pi}{6} \sum_{j=1}^N q_j \,
\delta^{(2)}(x-x_j)/\sqrt{g(x_j)} \ . \ee

As discussed, e.g., in \cite{David:89,BNT:2000,BowTra:01a},
minimizing (\ref{hexen}) subject to a fixed arrangement of $N$
defects, leads to \bea \label{defen} E &=& \frac{K_A}{2} \int
d^2x\sqrt{g(x)}\int d^2y\sqrt{g(y)}
\nonumber\\
&\times& \left[s(x)-K(x)\right]G(x,y)\left[s(y)-K(y)\right]
\nonumber\\
&+& \frac{\kappa}{2}\int d^2x\sqrt{g(x)}\,H^2(x) \ , \eea where
$K(x)$ is the Gaussian curvature and $G(x,y)$ is the inverse
Laplacian on the torus. Up to subtractions which eliminate ``zero
modes" (see Sec.~\ref{SECT__IDT}), $G(x,y)$ obeys
\bea\label{lapong} \nabla_x^2 G(x,y) &=&
\frac{1}{\sqrt{g}}\partial_i\{\sqrt{g}g^{ij}\partial_j\}G(x,y)
\nonumber\\
&=& \delta^{(2)}(x-y)/\sqrt{g(x)} \ . \eea The first term of
(\ref{defen}) arises directly from (\ref{hexen}) and represents a
kind of two-dimensional electrostatics in curved space. As
discussed below, this electrostatics can lead to a lower energy
when positive and negative disclinations are placed at positions
on the surface which approximately match the local Gaussian
curvature. To account for the bending energy of the surface, we
have added the second (Helfrich) term, where the bending rigidity
coupling is $\kappa \approx (1{-}10)\,{\rm k}_BT$ for lipid
bilayers and $H(x)$ is the mean curvature. Defect core energies,
which depend on short distance physics not accounted for in this
continuum approach, will be added later.

\begin{figure}[ht]
\includegraphics[width=2in]{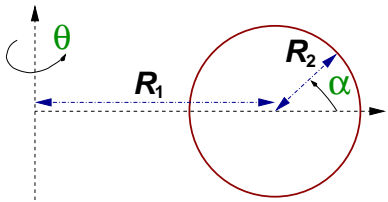}
\caption{Coordinates $(\alpha,\theta)$ defining the torus.}
\label{fig__CO}
\end{figure}

Although a representation of the physics of geometrical
frustration like (\ref{hexen}) is possible {\em locally} on any
smooth surface, coordinates for genus one surfaces (the torus) can
be found which admit such a representation {\em globally}. One
such coordinate system is shown in Fig.\ref{fig__CO}, where a
point on the torus is specified by \be\label{globcoords}
\vec{R}(\alpha,\theta) = \left(
\begin{array}{c}
\left[ R_1 + R_2 \cos \alpha \right] \cos \theta \\
\left[ R_1 + R_2 \cos \alpha \right] \sin \theta \\
R_2 \sin \alpha \end{array} \right) \ee

\noindent Here $R_2 < R_1$ so that the torus does not intersect
itself. The Gaussian curvature associated with (\ref{globcoords})
is a function of $\alpha$ only, \be\label{gausscurv} K(\alpha) =
\frac{\cos \alpha}{R_1 R_2 \left[1 + \frac{R_2}{R_1} \cos \alpha
\right]} \ \ . \ee Note that $K$ is positive on the outside wall
of the torus $(-\frac{\pi}{2} < \alpha < \frac{\pi}{2})$ and
negative on the inside wall $\left( \frac{\pi}{2} < \alpha <
\frac{3\pi}{2}\right)$. Although the $(\alpha,\theta)$ coordinate
system has a clear physical interpretation, most of our results
are obtained using the {\em conformal} coordinates introduced in
the Appendix~\cite{MF:1967}. Upon replacing $\alpha$ by a new
angular variable $\varphi \, (0 \leq \varphi \leq 2\pi)$ defined
by \be \cos \varphi = \frac{R_1 \cos \alpha + R_2}{R_1 + R_2 \cos
\alpha} \ , \ee one obtains a locally {\em flat} metric, which
greatly simplifies the calculations.

\begin{figure}[hb]
\includegraphics[width=3in]{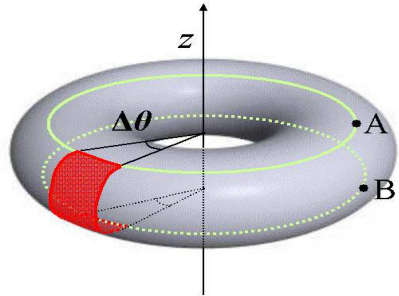}
\caption{Patch of positive Gaussian curvature on the torus
(shaded) subtending an azimuthal angle $\Delta\theta$. The surface
of the torus is divided into regions of positive and negative
Gaussian curvature by the circles labelled A and B, on which the
Gaussian curvature vanishes (after D. Hilbert and S. Cohn-Vossen,
{\em Geometry and the Imagination} (Chelsea, New York, 1952).}
\label{pospatch}
\end{figure}

The ``electrostatic" energy term in Eq.(\ref{defen}) favors
approximately charge neutral configurations, with discrete
disclination charges cancelling the smeared out Gaussian
``curvature charge". Although the full calculation (with core
energies taken into account) is subtle (see Sec.\ref{SECT__HTV}),
it is interesting to estimate how many additional disclinations
might be accommodated on a torus. In the torus shown in
Fig.\ref{pospatch}, the solid and dashed circles divide the
surface into regions of positive and negative Gaussian curvature.
Consider a wedge of angular width $\Delta\theta$ on the outside
wall of positive Gaussian curvature. The net ``curvature charge"
associated with this region is \be \Delta K =
\int^{\Delta\theta}_0
d\theta\int^{\pi/2}_{-\pi/2}d\alpha\,\sqrt{g}\,K(\alpha) =
2(\Delta\theta) \ , \ee where we have used Eq.(\ref{gausscurv})
and $ \sqrt{g} = R_1R_2\left(1 + \frac{R_2}{R_1} \cos
\alpha\right) $. Upon equating $\Delta K$ to $2\pi/6$, the charge
of a single disclination, we find that $\Delta\theta=2\pi/12$,
independently of $R_1$ and $R_2$. Thus $2\pi/\Delta\theta = 12$
positive disclinations would be required to completely compensate
the curvature of the outer wall. Similarly, 12 negative
disclinations would be required to completely compensate the
negative curvature of the inner wall. This simple argument
neglects core energies and interactions between disclinations,
effects which will cause the preferred number of defect pairs to
be less than 12.

The thermodynamic limit for torus dimensions $R_1$ and $R_2$
corresponds to the limits $R_1/a_0 \rightarrow \infty$ and
$R_2/a_0 \rightarrow \infty$, with the aspect ratio $r=R_1/R_2$
fixed, where $a_0$ is a microscopic length scale such as the
particle spacing. Upon optimizing $r$ with a defect-free hexatic
texture on the torus, one finds that a nonzero hexatic stiffness
constant  $K_A$ pulls $r$ above the Clifford torus value
$r=\sqrt{2}$ appropriate for liquid torii \cite{LuPr:92}, so that
the resulting shape looks more like a bicycle tire (see
Sec.~\ref{SECT__HTV}{-}A). Our results for defect energies are
presented for fixed $r$ and a given number of $M \sim
R_1R_2/{a_0}^2$ of microscopic degrees of freedom. It would be
straightforward, however, to use the methods described here to
optimize over {\em both} $r$ and possible defect configurations
for fixed $M$.

Although we find that disclinations are always unfavorable in the
``thermodynamic limit" of large $M$, the critical value $M=M_c$
required to suppress them in the ground state is surprisingly
large. Indeed, for $r=\sqrt{2}$, this number is of order
$10^{11}!$ (see Fig.\ref{fig__MMax}). As $r$ approaches 1 from
above, i.e., in the limit of ``fat" torii, we find that $M_c$
exhibits a remarkably strong divergence [see
Eq.(\ref{masterformula})], \be \label{hugecritexp} M_c \sim
\frac{1}{(r-1)^{12}} \ , \ee as treated in
Sec.~\ref{SECT__HTV}{-}C. Because $M_c$ is so large, the toroidal
vesicles of Ref.~\cite{MuBe:91} would be quite likely to have
disclinations present in the ground state if hexatic order were
present. Indeed, for $R_1=5\mu$ (roughly the size of a red blood
cell), $R_2=rR_1 \approx 1.4R_1$ and $a_0=20\AA$ (typical lipid
spacing in vesicles), we have (see Sec.~\ref{SECT__HTV}{-}B)
$M=\frac{8\pi^2}{\sqrt{3}}\frac{R_1R_2}{a_0^2} \approx 4 \times
10^8$ which is much less than the critical value $M_c \sim
10^{11}$ required to suppress disclinations. As discussed above,
the interaction between hexatic order and the Gaussian curvature
of toroidal vesicles leads to $r > \sqrt{2}$ and a smaller value
of $M_c$. Vesicles with lipids in a liquid state could provide,
however, a toroidal template with $r = \sqrt{2}$ for adsorbed
colloidal particles, similar to the spherical ``colloidosomes"
studied by Dinsmore et. al. \cite{DHNMBW:2002}. It may be possible
to use polymerizable lipids to permanently fix the template aspect
ratio at $r=\sqrt{2}$. It would be quite interesting to study
(with, say, confocal microscopy) both hexatic and crystalline
order in colloidal particle arrays \cite{CMurray:1992} adsorbed on
such a template, as has already been done for colloids on
spherical water droplets in oil \cite{Scars:2003}. The colloid
radius would play the role of a microscopic scale $a_0 \ll
R_1,R_2$ in this case. Disclination defects in a crystalline
ground state might well be accompanied by grain boundaries
\cite{BNT:2000}.

Although we focus here on hexatic order in toroidal geometries,
similar results should apply to XY-like models, as would be
appropriate for vesicles composed of lipid bilayers with tilted
molecules \cite{SSSC:88,SSSPC:90}. Our results are relevant as
well to two-fold nematic order on a toroidal template. In both
cases, we expect qualitatively similar phenomenon, such as defects
in the ground state, unless the total number of degrees of freedom
exceeds a critical value. A precise equivalence is possible in the
one Frank constant approximation \cite{deGPro:93}. As discussed in
Sec.~\ref{SECT__HTV}, defects in the ground state are more likely
for fat torii in the case of nematic (and hexatic) order.
Interesting results related to those here have recently been
obtained for ``corrugated" topographies, which are flat at
infinity and for which there is also no topological necessity for
defects in the ground state \cite{VN:2003}. Specifically, it has
been shown that defect pairs lower the energy of a hexatic phase
draped over a Gaussian ``bump" for sufficiently large ratio of
height to width. In this case, defects remain an important feature
of the ground state even when the number of degrees of freedom $M$
tends to infinity.

The organization of the paper is as follows: an analogy to an
electrostatic problem, aimed at providing a more intuitive picture
of the physics of curvature-induced defect unbinding, is
introduced in Sec.~\ref{SECT__EL}. In Sec.~\ref{SECT__IDT} the
interaction among defects is worked out in detail for toroidal
topology. Predictions for the total number of defects are provided
in Sec.~\ref{SECT__HTV}. The effects of both temperature and shape
fluctuations are discussed in Sec.~\ref{SECT__TEMP}.

\section{Electrostatic Analogy}
\label{SECT__EL}

\begin{figure}[hb]
\includegraphics[width=2in]{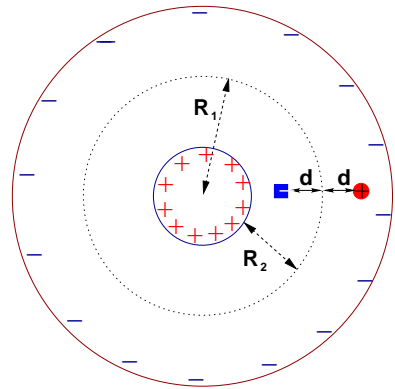}
\caption{A plus (filled circle) and a minus (filled square) pair
of charges between the plates of a two-dimensional circular
capacitor is analogous to a plus-minus pair of disclinations on a
torus with the identifications given.} \label{fig__el}
\end{figure}

\begin{figure}[ht]
\includegraphics[width=2.5in]{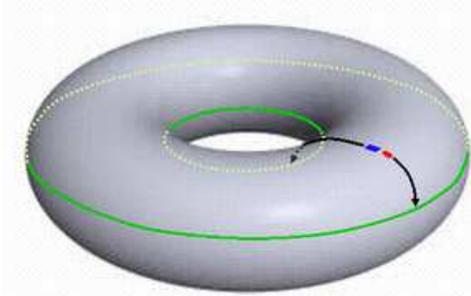}
\caption{Illustration of the calculation discussed in the text: a
plus (filled circle) and a minus (filled square) form a
disclination dipole on one of the two circles of zero Gaussian
curvature. They are then pulled apart until they reach the maximum
curvature line (plus) and minimum curvature line (minus).}
\label{fig__Ill1}
\end{figure}

It is useful to illustrate the physics of defects on the torus
with a simple electrostatic analogy, illustrated in
Fig.~\ref{fig__el}. A positive and negative charge are placed
between the plates of a capacitor with circular cross section. The
plus-minus pair are the analogs of a plus-minus disclination
dipole initially located on a zero curvature line of the torus
(see Fig.~\ref{fig__Ill1}). We assume here that the charges are
extended over a core radius $a_0$, which plays the role of a
minimum separation. We expect that this core radius is related to
the mean particle separation on the torus. The net charge $+Q$ and
$-Q$ on the capacitor plates represents the Gaussian curvature in
Eq.(\ref{defen}), integrated over the regions of the torus where
it is positive and negative, respectively.  The linear charge
density on the plates of the capacitor is thus given by
$\rho_{\pm}=Q/L_{\pm}$, where $L_{\pm}=2\pi(R_1 \mp R_2)$. The
lengths $R_1$ and $R_2$ for this capacitor correspond to the torus
radii defined in Fig.~\ref{fig__CO}.

The competition between the work done by the electric field of the
capacitor (representing regions of $+/-$ Gaussian curvature on the
torus) and the attraction of opposite sign charges will dictate
whether the disclinations separate or remain tightly bound at
separation $a$. Since the energy of a configuration with excess
bound charges exceeds that for no excess charges by an amount of
order two core energies, this criterion determines whether a
separated plus-minus pair is present in the ground state.

In this two-dimensional geometry, the electrostatic energy of the
two charges with separation $2d$ and charge $e$ is given by
\be\label{en_charges} {\cal E}_1= e^2 \ln(d/a_0)+ 2E_{c} \ , \ee
where $E_c$ represents a self-energy of an isolated charge,
corresponding to the core energy of a disclination. There is an
additional electrostatic force pulling the charges to the
capacitor plates, which leads to an additional energy
\be\label{en_capacitor} {\cal E}_2= Q e
\ln\left(\frac{R_1-d}{R_1+d}\right) \ . \ee  The total energy is
then \be\label{en_tot} {\cal E}=2\pi e^2 {\mathcal Q}-Qe {\cal L}
+ 2E_c \ , \ee where \bea
{\cal Q}&=&\frac{1}{2\pi}\ln(d/a_0) \ , \nonumber\\
{\cal L}&=&\ln\left(\frac{R_1+d}{R_1-d}\right) \ . \eea The
functions ${\mathcal Q}$ and ${\mathcal L}$ have a similar form to
those we find in the exact calculation on the torus (see
Eq.(\ref{I_functions})).

Eq.(\ref{en_tot}) illustrates very clearly the appearance of two
preferred locations; these being $d=a_0$ (where ${\cal E} \approx
2E_c$) and $d=R_2$ $\left( {\rm where} \, {\cal E} = 2E_c +
e^2\ln(R_2/a_0) -
eQ\ln\left[\frac{R_1+R_2}{R_1-R_2}\right]\right)$. The relative
strength of the first two terms will determine the preferred
location of the charge dipole. By taking $a_0$ small enough, for
$R_1 > R_2$, the attractive charge-charge term will dominate and
defects will not be favored. For any finite $a_0$, on the other
hand, defect unbinding will be favored in the limit $R_2
\rightarrow R_1^-$. This is precisely what we find, up to
numerical constants, in our treatment of disclination unbinding on
the torus.

\section{Interacting Defects on Curved Surfaces}
\label{SECT__IDT}

\subsection{Green's function on the Torus}

For arbitrary coordinates $\vec{{\bf x}} = (x_1,x_2)$ on the
torus, the key object in the energy Eq.~(\ref{defen}) is the
inverse Laplacian, which is the solution $G(\vec{{\bf
x}},\vec{{\bf x^{\prime}}})$ to the equation \be\label{inv_Lap}
\Delta G({\bf x},{\bf x^{\prime}})=\delta({\bf x},{\bf
x^{\prime}})-\frac{1}{A} \ , \ee where the Laplacian is defined as
in Eq.(\ref{lapong}), $\delta({\bf x},{\bf
x^{\prime}})=\frac{1}{\sqrt{g}}\delta({\bf x}-{\bf x^{\prime}})$,
and $A$ is the area of the torus, \be\label{area} A\equiv\int
d^2{\bf x}\,\sqrt{g}= 4\pi^2 R_1 R_2 \ . \ee A constant is
subtracted from the $\delta$-function to eliminate a ``zero mode"
which changes the area of the torus. As shown in the Appendix, the
metric of a torus, with a suitable change of coordinates, is flat
modulo an overall conformal factor [see Eq.(\ref{torus_rr_mm})].
The inverse Laplacian may be computed by considering the flat
metric (where the conformal factor is identically one) and its
associated inverse Laplacian $\widehat{G}({\bf x},{\bf
x^{\prime}})$, from the formula \bea\label{Green}&G({\bf x},{\bf
x^{\prime}})&=\widehat{G}({\bf
x},{\bf x^{\prime}}) \nonumber\\
&-& \int\frac{d^2{\bf y}}{A}\sqrt{g(\bf
y)}\,\left[\widehat{G}({\bf x},{\bf y})
+ \widehat{G}({\bf y},{\bf x^{\prime}})\right]\nonumber\\
&+&\int\frac{d^2{\bf y}}{A}\int\frac{d^2{\bf y^{\prime}}}{A}
\sqrt{g({\bf y})}\sqrt{g({\bf y^{\prime}})} \widehat{G}({\bf
y},{\bf y^{\prime}})  \eea As can be checked straightforwardly,
$G({\bf x},{\bf x^{\prime}})$ solves Eq.~(\ref{inv_Lap}) as well
as satisfying \bea\label{add_cond} \int d^2{\bf x} \sqrt{g({\bf
x})} \,G({\bf x},{\bf x'})&=&0
\\\nonumber
\int d^2{\bf x^{\prime}} \sqrt{g({\bf x^{\prime}})}\,G({\bf
x},{\bf x^{\prime}})&=&0 \ . \ \eea Thus Eq.(\ref{Green}) is
indeed the inverse Laplacian, where the conditions
(\ref{add_cond}) ensure overall ``charge neutrality" for any
disclinations present on the torus.

The coordinates for the torus are shown in Fig.~\ref{fig__CO}.
Upon making the change of variables $\alpha \rightarrow \varphi$
via $\cos \alpha \equiv \frac{R_1 \cos \varphi - R_2}{R_1 - R_2
\cos \varphi}$, described in the Appendix, the inverse Laplacian
$\widehat{G}({\bf x},{\bf x^{\prime}})$ in conformal coordinates
can be computed by first solving \bea\label{inv_lap2}
-(\sinh\rho\,\partial^2_{\theta}+\frac{1}{\sinh\rho}\partial_{\varphi}^2)
\widehat{G}(\theta,\varphi|\theta',\varphi')&=&
\nonumber\\\
\delta(\theta-\theta',\varphi-\varphi')- \frac{1}{(2\pi)^2} && \ ,
\eea where $\sinh\rho = \sqrt{r^2-1}$ and $r$ is the aspect ratio
$R_1/R_2$ of the torus. The straight-forward solution
\bea\label{naive_sol} {\widehat
G}^{(0)}(\theta,\varphi|\theta',\varphi')&=& \nonumber
\\ -\frac{1}{4\pi}\ln\left[ (\sinh
\rho){(\theta-\theta^{\prime})}^2+\frac{1}{\sinh\rho}{(\varphi-\varphi^{\prime})}^2
\right ]&,& \eea satisfies $\Delta {\widehat
G}^{(0)}=\delta(\theta-\theta',\varphi-\varphi')$, but is not
periodic, i.e.,  \bea\label{pp_cond} {\widehat
G}^{(0)}(\theta+2\pi k,\varphi|\theta',\varphi')&\not=& {\widehat
G}^{(0)}(\theta,\varphi+2\pi n|\theta',\varphi')
\nonumber\\
&\not=&{\widehat G}^{(0)}(\theta,\varphi|\theta',\varphi') \ ,
\eea for arbitrary integers $k$ and $n$. In addition the Laplacian
acting on ${\widehat G}^{(0)}(\theta,\varphi|\theta',\varphi')$
fails to give the constant term on the right hand side of
Eq.(\ref{inv_lap2}). Both these deficiencies are remedied by
defining \bea\label{per_sol} {\widehat
G}(\theta,\varphi|\theta',\varphi')&=&-\frac{1}{2\pi}\sum_{k,n}\ln[
\sinh\rho(\theta + 2\pi k)^2 \nonumber\\
&+&\frac{1}{\sinh \rho}(\varphi + 2\pi n)^2]  + {\mathcal C}(\rho)
\ , \eea where the constant ${\mathcal C}(\rho)$ is determined by
imposing that the inverse Laplacian exhibits the correct
short-distance singularity (following from Eq.(\ref{naive_sol}))
in the limit $\theta\rightarrow\theta'$ and
$\varphi\rightarrow\varphi'$. Standard analytical techniques
\cite{Polchinski:1998,Mumford} allow one to perform the sum
indicated in Eq.~(\ref{per_sol}), giving \bea\label{G_true}
{\widehat
G}(\theta,\varphi|\theta',\varphi')&=&-\frac{1}{4\pi}\ln\left[\frac{\sinh
\rho |\vartheta_1(
\frac{\theta-\theta'+\tau(\varphi-\varphi')}{2\pi},\tau)|^2}
{4\pi^2|\eta(\tau)|^6}\right]\nonumber\\
&+&\frac{1}{2\sinh\rho}\left(\frac{\varphi-\varphi'}{2\pi}\right)^2
\ , \eea where $\tau=\frac{i}{\sinh \rho}$. The functions
$\vartheta_1$ and $\eta$ are the Theta function and Dedekind
$\eta$ function, respectively, defined by
\cite{WhitWat}\be\label{theta1}
\vartheta_1(\nu,\tau)=-i\sum_{n=-\infty}^{n=+\infty}(-1)^n
e^{i\pi\tau(n-1/2)^2}e^{2\pi i \nu (n-1/2)}  \ee and \be
\label{theta2} \eta(\tau)=e^{2\pi i
\tau/24}\prod_{n=1}^{\infty}(1-e^{2\pi i n \tau}) \ . \ee

 The inverse Laplacian of Eq.(\ref{Green}) thus becomes
 \bea\label{green_torus}
G(\theta,\varphi|\theta',\varphi')={\widehat
G}(\theta,\varphi|\theta',\varphi') &-&
\nonumber\\
\frac{2}{(2\pi)^2} \left(
\frac{1}{\sinh\rho}\sum_{n=1}^{\infty}\frac{e^{-n\rho}}{n^2}[\cos(n\varphi)+\cos(n\varphi')]
\right. && \nonumber\\\nonumber \left.
+\frac{1}{\cosh\rho}\sum_{n=1}^{\infty}\frac{e^{-n\rho}}{n}[\cos(n\varphi)+\cos(n\varphi')]
\right)&+&
\\\nonumber
+\frac{2}{(2\pi)^2}\left(
\frac{1}{\sinh\rho}\sum_{n=1}^{\infty}\frac{e^{-2n\rho}}{n^2}+
\right. &&
\nonumber\\
\left. \frac{2}{\cosh\rho}\sum_{n=1}^{\infty}\frac{e^{-2n\rho}}{n}
\frac{\tanh\rho}{\cosh\rho}\sum_{m=1}^{\infty}e^{-2m\rho}\right)
&\ ,& \eea which can then be used to evaluate Eq.~(\ref{defen}).

\subsection{Energetics of Defects on a Torus}

The total energy Eq.(\ref{defen}) also contains a bending rigidity
term which, for a torus with aspect ratio $r=R_1/R_2$, is
\be\label{int_pos} {\cal
E}^{\kappa}=\frac{2\pi^2r^2}{\sqrt{r^2-1}}\kappa \ .
 \ee
If we were to minimize this term alone, we would find
$r=\sqrt{2}$, the so-called Clifford torus. The Gauss-Bonnet
theorem for a torus reads \be\label{GB_TH} \int d^2{\bf x}
\sqrt{g}\,K({\bf x})=0 \ , \ee and, with our choice of Green's
functions, Eq.(\ref{hexen}) insures that the sum of disclination
``charges" $q_i$ satisfies \be \sum_{i=1}^N q_i=0  \ee so that, as
previously noted, no defects are required topologically. The
defect charges here take on the values $q_i=\pm 1$,$\pm 2$,
$\ldots$, with $q_i=\pm 1$ corresponding to the elementary defects
with rotations $\pm 2\pi/6$ in the hexatic order parameter.

With the Green's function in hand, the hexatic energy of a set of
disclination charges on a torus (the first part of
Eq.(\ref{hexen})) can thus be written explicitly:
\bea\label{Ising_en} {\mathcal E} &=& \frac{\pi^2K_A}{18}
\sum_{i{\neq}j}^N q_i q_j {\mathcal Q}({\bf x}_i,{\bf x}_j)
-\frac{\pi K_A}{3}\sum_{i=1}^Nq_i {\mathcal L}({\bf x}_i)
\nonumber\\ &+& {\mathcal D} + \left(\sum_{i=1}^{N}q_i^2\right)E_c
\ , \eea where the defects interact with each other according to
\bea\label{I_functions} {\mathcal Q}({\bf x}_i,{\bf x}_j)&=&
-\frac{1}{4\pi}\ln\left(\frac{|\vartheta_1(\frac{\theta_i-\theta_j}{2\pi}
+ \frac{i(\varphi_i-\varphi_j)}{2\pi\sinh\rho},
\frac{i}{\sinh\rho})|^2}{4\pi^2|\eta(\frac{i}{\sinh\rho})|^6}\right)
\nonumber \\ & + &
\frac{1}{2\sinh\rho}\left(\frac{\varphi_i-\varphi_j}{2\pi}\right)^2
\ , \eea and interact with the background Gaussian curvature
``charge" according to \be\label{bgcurvint} {\mathcal L}({\bf
x})=\ln\left(\frac{1}{\cosh\rho-\cos\varphi}\right) \ . \ee The
``spin wave" part of the frustrated hexatic energy
\be\label{spinwave} {\mathcal
D}=\frac{1}{2}K_A(2\pi)^2e^{-\rho}=\frac{2{\pi}^2K_A}{r+\sqrt{r^2-1}}
\ee  is present even without defects and supplements the bending
rigidity term Eq.~(\ref{int_pos}). The core energy term in
Eq.~(\ref{Ising_en}) will be considered in more detail in
Sec.~\ref{SECT__HTV}.

\subsection{Energetics of Defects on a Sphere}

It is instructive to compare our results for toroidal vesicles
with the corresponding results for spherical vesicles. The
interaction potential (\ref{I_functions}) for a sphere is
\be\label{I_functions_sp} {\mathcal Q}({\bf x}_i,{\bf x}_j)=
-\frac{1}{4\pi}\ln\left(\frac{1-\cos \beta_{ij}}{2}\right) \ , \ee
where $\beta_{ij}$ is the geodesic distance, for a sphere of unit
radius, between points ${\bf x}_i$ and ${\bf x}_j$. The function
${\mathcal L}(\bf{x})$ and the constant $\cal{D}$ can both be set
to zero on the sphere.

The Gauss-Bonnet theorem for spherical topology reads
\be\label{GB_SP} \int d^2{\bf x} \sqrt{g}\,K({\bf x})=4\pi \ , \ee
yielding the constraint \be \sum_{i=1}^N q_i=12  \ . \ee

\section{Ground States of Hexatic Toroidal Vesicles}\label{SECT__HTV}

Before presenting a detailed analysis of the implications of the
Hamiltonian (\ref{Ising_en}) for defects on a torus we must
confront the issue of core energies. The short-distance behavior
of the defect potential ${\cal Q}$ (Eq.(\ref{I_functions}))
implies that a plus-minus pair of disclinations located as in
Fig.\ref{fig__Ill1} on a circle of fixed azimuthal angle $\theta$
and geodesic separation $2d$ have an energy \be\label{short_dist}
E=\frac{K_A \pi}{18}\left[\ln(R_2/a_0) + V_I(d/R_2)\right]+2E_{c}
, \ee where, after absorbing various constants, $V_I(d/R_2)$ may
be viewed as the interaction potential between the two
disclinations and the term $E_{c}$ (reflecting short-distance
physics) has been added by hand. For $a_0 \ll 2d \ll R_2$,
$V_I(d/R_2) \simeq \ln(2d/R_2)$. The torus radius $R_2$ therefore
drops out in this limit and we recover the result
$E={K_A\pi}/18\ln(2d/a_0) + 2E_c$ for a disclination pair in flat
space.  The disclination core radius $a_0$ is the distance (of
order the spacing between molecules) at which continuum elasticity
breaks down. We will henceforth assume that
\be\label{Core_functions} E_{c} \approx  c K_A \ , \ee where $c$
is a dimensionless parameter characterizing the size of the
disclination core energy. In our numerical calculations we take $c
= 0.1$.

If the torus is coated with $M$ particles, an effective average
particle spacing $a_P$ (assuming a local triangular lattice for
simplicity) may be defined from the area per particle
\be\label{Latt_const} \frac{\sqrt
3}{2}a_{P}^2=\frac{4\pi^2R_1R_2}{M}=\frac{4\pi^2 R^2_2\,r}{M} \ .
\ee We shall assume that $a_0 \approx a_P$, so
Eq.(\ref{Latt_const}) relates the total number of particles $M$ to
the minimum plus-minus pair separation appearing in
Eq.(\ref{short_dist}).

Upon including both the bending rigidity and the core energies,
the total energy for $N$ charge $\pm 1$ defects on a rigid,
undeformed torus takes the form \bea\label{Ising_tot} &{\cal E}&=N
E_{c}+\frac{K_A}{2}\left( \frac{2 \pi^2}{9} \sum_{i=1}^N
\sum_{j>i}^N q_i q_j {\cal Q}({\bf x}_i,{\bf x}_j) \right.
\nonumber\\
&-&\left. \frac{2\pi}{3}\sum_{i=1}^Nq_i {\cal L}({\bf x}_i)+
\frac{4\pi^2}{r + \sqrt{r^2-1}} \right)+\kappa \frac{2 \pi^2
r^2}{\sqrt{r^2-1}} \eea ${\mathcal E}$ depends on the lattice
spacing $a_0$ through the core energy Eq.~(\ref{Core_functions})
as well as the constraint that the defect spacing cannot be
smaller than $a_P$. In our calculations the latter constraint is
accounted for by forbidding any two defects from approaching
within the distance $a_P$ or, alternatively, by assuming that any
two defects closer than $a_P$ merge into a single defect with
total charge the sum of the individual charges of the two defects.

Only the potential (${\cal Q}$) and curvature (${\cal L}$) terms
depend explicitly on the defect positions and charges. For fixed
$R_1$ and $R_2$, therefore, the total number of defects in the
ground state is independent of both the bending rigidity and the
constant ${\cal D}$ in Eq.(\ref{spinwave}).

\begin{figure}[ht]
\includegraphics[width=3in]{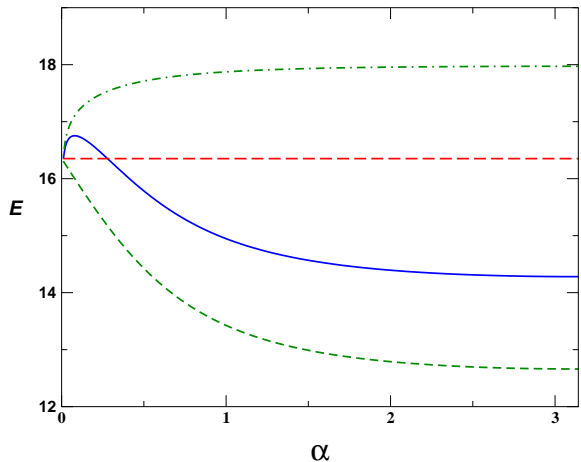}
\caption{Various contributions to the energy (in units of $K_A/2$)
with aspect ratio $r=\sqrt{2}$ for a disclination dipole separated
an angular distance $\alpha$ along the path shown in
Fig.~\ref{fig__Ill1}. The long dashed line is the energy in the
absence of defects. The dashed-dotted line is the defect-defect
interaction as a function of the separation of the charges. The
dashed line is the curvature-defect interaction energy as a
function of separation, and the continuous line is the total
energy. The core energy contribution, computed from
Eq.~\ref{Core_functions}, is very small on this scale.The bending
energy is subtracted for clarity.} \label{fig__psi_all}
\end{figure}

The defect-defect interaction is determined through the ${\cal
Q}$-function defined in Eq.~(\ref{I_functions}). For two opposite
sign defects the energy is attractive for all separations at
constant $\theta$, as shown in Fig.~\ref{fig__psi_all}. If only
this term were present, the attraction would bring both charges as
close as possible, binding all disclinations into dipoles which
have a higher energy than a defect-free configuration. Thus if no
other terms were present, the ground state would be defect free.

The defect-curvature interaction ${\cal L}$ favors the appearance
of additional defects. This term acts like an electric field
pulling the positive/negative disclinations into regions of
positive/negative Gaussian curvature, respectively, similar to the
electrostatic analogy discussed in Sec.\ref{SECT__EL}. As shown in
Fig.~\ref{fig__psi_all}, if this were the only term present,
isolated plus and minus disclination charges are always
energetically favored, with the lowest energy arising when they
are located at the regions of absolute maximum Gaussian curvature.

The total energy is a competition between defect attraction and
curvature-induced unbinding, as shown in Fig.~\ref{fig__psi_all},
where the potential shows a double well, corresponding to the
attractive defect-defect interaction minimum and the attractive
Gaussian curvature minimum. In Fig.~\ref{fig__psi_all}, the
curvature field dominates and additional defects appear in the
ground state, but the general result for whether defects should or
should not be expected is a function of the core energy $E_c$, the
hexatic stiffness $K_A$, the torus aspect ratio $r = R_1/R_2$, and
the ratio $R_2/a_P$ of macroscopic to microscopic cutoff.

\subsection{Defect-free hexatic toroidal vesicles}

In the absence of defects the total energy following from
Eq.~(\ref{Ising_en}) is \be\label{rr_vv} E=\frac{2\pi^2
K_A}{r+\sqrt{r^2-1}}+\kappa \frac{2\pi^2r^2}{\sqrt{r^2-1}} \, \ ,
\ee a result first obtained in \cite{LuPr:92}. The optimal value
of $r$ resulting from minimizing this energy is the solution of
\be\label{Eq_sol}
(r^2-1)^{3/2}-r(r^2-1)+\frac{\kappa}{K_A}r(r^2-2)=0 , \ee which
read in the limit of large and small hexatic stiffness
\cite{LuPr:92},
\bea\label{sol_two} r&=& \sqrt{2} \  ,  \  K_A <<
\kappa
\nonumber\\
r&=& \sqrt{\frac{K_A}{2\kappa}} \  , \  K_A >> \kappa \ . \eea If
the hexatic stiffness is much smaller than the bending rigidity,
the Clifford torus ($r=\sqrt{2}$) is the optimal geometry. If, on
the other hand, the hexatic stiffness dominates, then a thin
torus, similar to a bicycle tire, is optimal. This picture changes
when defects are included.

\subsection{Ground states of defective hexatic toroidal vesicles}

The general ground state for arbitrary aspect ratio may be
determined numerically using Eq.~(\ref{Ising_tot}). For
simplicity, we compare the energies of configurations with and
without defects for a fixed aspect ratio $r$.

We first performed the following calculation: a set of $N$ unit
charge disclinations ($N/2$ positive and $N/2$ negative) are
placed in opposite-sign pairs on a circle of zero Gaussian
curvature. Each pair is then pulled apart at constant azimuthal
angle $\theta$ until the plus(minus) disclinations reach the
outer(inner) rim of the torus respectively. This is illustrated in
Fig.~\ref{fig__Ill1} for the simplest case $N=2$.

\vspace{0.2cm}

\begin{figure}[hb]
\includegraphics[width=2in]{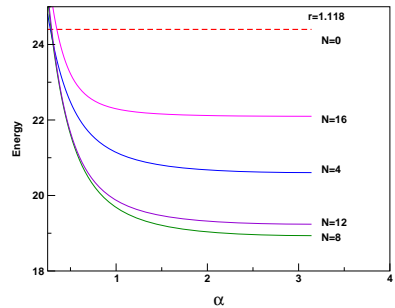}
\caption{The total energy (in units of $K_A/2$) for aspect ratio
$r=\sqrt{5/4}$ for varying numbers of defects and $a_P=R_2/4$. The
bending energy at fixed $r$ is subtracted off. The disclination
core energy is taken to be $0.1K_A$, which is $0.2$ in the above
units (corresponding to $c=0.1$).} \label{fig__no_core1a}
\end{figure}

\begin{figure}[ht]
\includegraphics[width=2in]{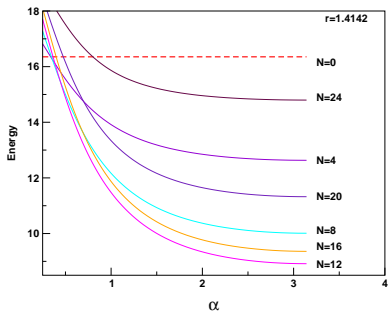}
\caption{The total energy, as in Fig.~\ref{fig__no_core1a}, but
for the Clifford torus which has aspect ratio $r=\sqrt{2}$.}
\label{fig__no_core1b}
\end{figure}

\begin{figure}[hb]
\includegraphics[width=2in]{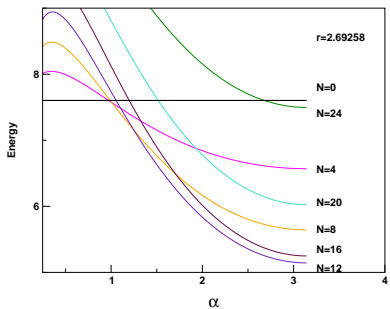}
\caption{The total energy, as in Fig.~\ref{fig__no_core1a}, but
for aspect ratio $r=2.6926$.} \label{fig__no_core1c}
\end{figure}
We discuss our results for two regimes of disclination core
energy. Core energies corresponding to $c$-coefficients less than
$1/10$ have almost no effect on the energy balance since the
elastic energy is already quite large. For definiteness the small
core energy regime will be illustrated for $c = \frac{1}{10}$. Our
results are shown in Figs.~\ref{fig__no_core1a},
\ref{fig__no_core1b} and \ref{fig__no_core1c} for the three aspect
ratios: $r=\sqrt{5/4}$, $r=\sqrt{2}$ (the Clifford torus) and
$r=2.69$. We have set $R_2/a_P=4$, so that the number of degrees
of freedom (see Eq.~(\ref{Latt_const})) on these torii is
$M=\frac{8\pi^2}{\sqrt{3}}r(\frac{R_2}{a_P})^2\simeq 800$, 1000
and 2000 respectively. In each case the addition of defects lowers
the energy. Note that the energy at maximum separation
($\alpha=\pi$) first decreases, and then increases with $N$. The
optimal number of defect pairs ($N/2$) is 5, 6 and 7 for Figs.
\ref{fig__no_core1a}, \ref{fig__no_core1b} and
\ref{fig__no_core1c} respectively. This number is less than the
naive estimate of 12 in the Introduction due to repulsive defect
interaction energies on the inner and outer walls of the torus.

\begin{figure}[ht]
\includegraphics[width=2in]{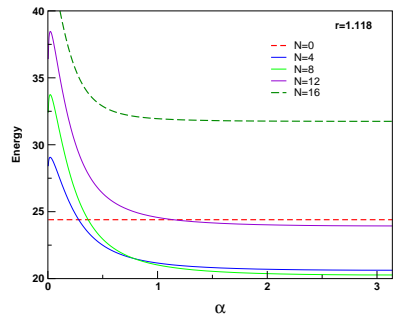}
\caption{The total energy (in units of $K_A/2$) for a torus with
aspect ratio $r=\sqrt{5/4}$ for varying numbers of defects with
$R_2/a_P=100/\pi$. The disclination core energy is taken to be
$K_A$ which is $2$ in the above units (corresponding to $c=1$).
The bending energy at fixed $r$ is subtracted off.}
\label{fig__yc_core1a}
\end{figure}

\begin{figure}[b]
\includegraphics[width=2in]{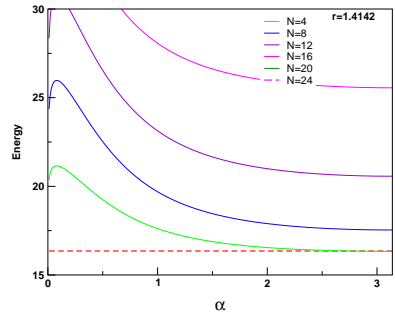}
\caption{The total energy, as in Fig.~\ref{fig__yc_core1a}, but
for aspect ratio $r=\sqrt{2}$.} \label{fig__yc_core1b}
\end{figure}

The typical situation for large core energies is illustrated for
$c = 1$. It is only for core energies of this order that we find
qualitatively different behavior from the small core energy
regime. Our results are plotted in Fig.~\ref{fig__yc_core1a}
($r=\sqrt{5/4}$) and Fig.~\ref{fig__yc_core1b} ($r=\sqrt{2}$).
Because $R_2/a_P= \frac{100}{\pi} \simeq 32$ in these plots, we
now have $M \approx 52,000$ and $M \approx 67,000$ respectively.
Even with such a large core energy, defects are present in the
ground state with a preferred number of pairs $N^*/2 \simeq 3$ for
Fig.~\ref{fig__yc_core1a} and a smaller number for
Fig.~\ref{fig__yc_core1b}. With this value of $R_2/a_P$ the torus
is defect free for larger $r$.

\begin{figure}[ht]
\includegraphics[width=3in]{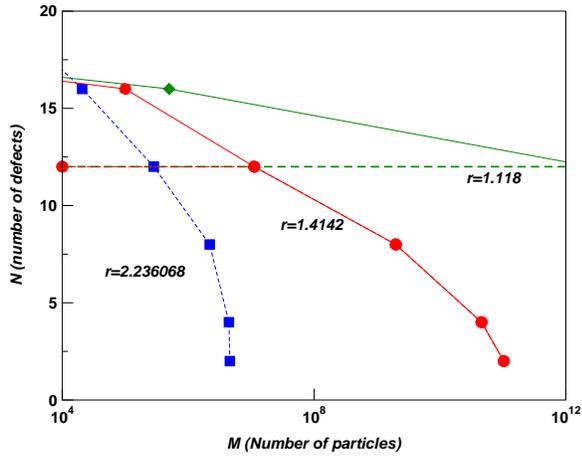}
\caption{The preferred number of defects as a function of the
total number of particles for three aspect ratios. The
disclination core energy is taken to be $0.1 K_A$. The dashed line
corresponds to the configuration in Fig.~\ref{fig__Ill1}.}
\label{fig__fM}
\end{figure}

Note that for fixed aspect ratio the preferred number of defects
drops for larger numbers of particles when the core energy is
large: compare Fig.~\ref{fig__yc_core1a} to
Fig.~\ref{fig__no_core1b}. To study this point further we have
determined the total number of defects in the ground state as a
function of the number of particles. Our results are shown in
Fig.~\ref{fig__fM}, using $M=\frac{8\pi^2}{\sqrt{3}}r(R_2/a_P)^2$
and assuming $a_0=a_P$. Although a torus always becomes defect
free in the thermodynamic limit $R_2/a_P \rightarrow \infty$ (with
$r=R_1/R_2$ fixed), torii with moderate aspect ratio only become
defect-free for a remarkably large number of particles, which may
be as large as $10^{11}$ for $r=\sqrt{2}$!

\begin{figure}[hb]
\includegraphics[width=2.5in]{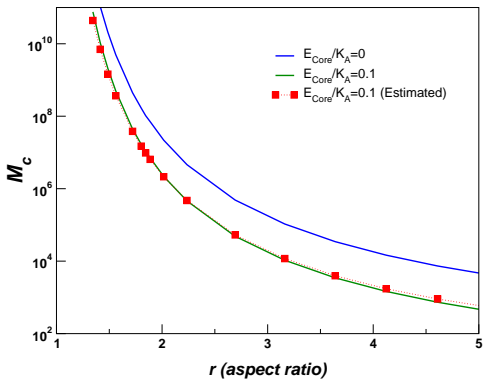}
\caption{The critical number of particles, above which defects are
no longer favored, as a function of the toroidal aspect ratio for
vanishing core energy (blue) and $c=0.1$ (green). The analytic
estimate for $c=0.1$ is plotted as solid red squares.}
\label{fig__MMax}
\end{figure}

To make the last point more transparent, the critical number of
particles $M_c$, above which defects are no longer favored, is
plotted as a function of the aspect ratio in Fig.~\ref{fig__MMax}.
As the aspect ratio $r=R_1/R_2 \rightarrow 1^+$, $M_{c}$ diverges,
suggesting that any torus will possess defects if sufficiently
fat. We provide an rough analytic argument along this line in the
next subsection.

Although fat torii (with $r \gtrsim 1$) tend to favor defects in
the ground state, the maximum number of defects favored for a
given aspect ratio is a subtle question. To see this, note that
the constrained minimization discussed above leads to a ring of
$N/2$ positive disclination charges on the outer wall of the
torus, and a smaller ring of $N/2$ negative charges on the inner
wall. As $r \rightarrow 1^{+}$, negative charges end up being very
close in the final configuration, with a considerable repulsive
energy cost.

\begin{figure}[ht]
\includegraphics[width=3in]{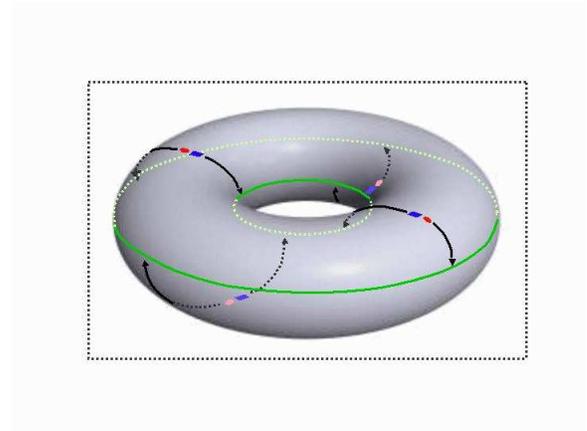}
\caption{Illustration of the calculation discussed in the text: a
plus disclination (filled circle) and a minus disclination (filled
square) form a defect dipole on one of the two zero-curvature
circles of the torus. They are then pulled apart in the direction
of the maximum curvature line (plus) and the minimum curvature
line (minus), respectively.} \label{fig__Ill2}
\end{figure}

\begin{figure}[ht]
\includegraphics[width=3in]{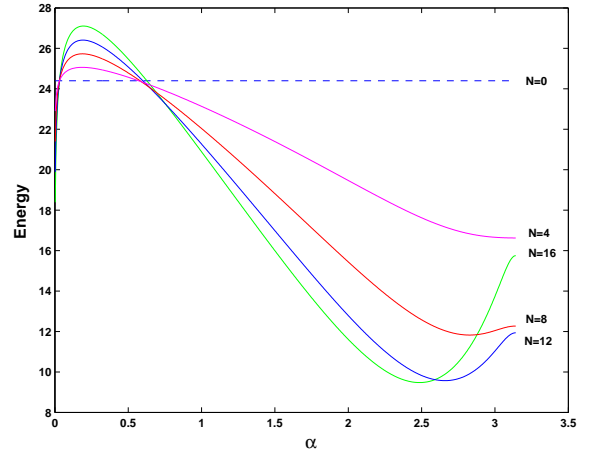}
\caption{Plot of the energy for the path described by
Fig.~\ref{fig__Ill2}. The disclination core energy is taken to be
$0.1K_A$. The aspect ratio is $r=\sqrt{5/4}$ and $R_2/a=50$. The
optimal number of defect pairs in the final ``buckled ring"
configuration is $N^*/2=7$ for this $M \approx 127,000$ particle
configuration.} \label{fig__Int}
\end{figure}

\begin{figure}[ht]
\includegraphics[width=3in]{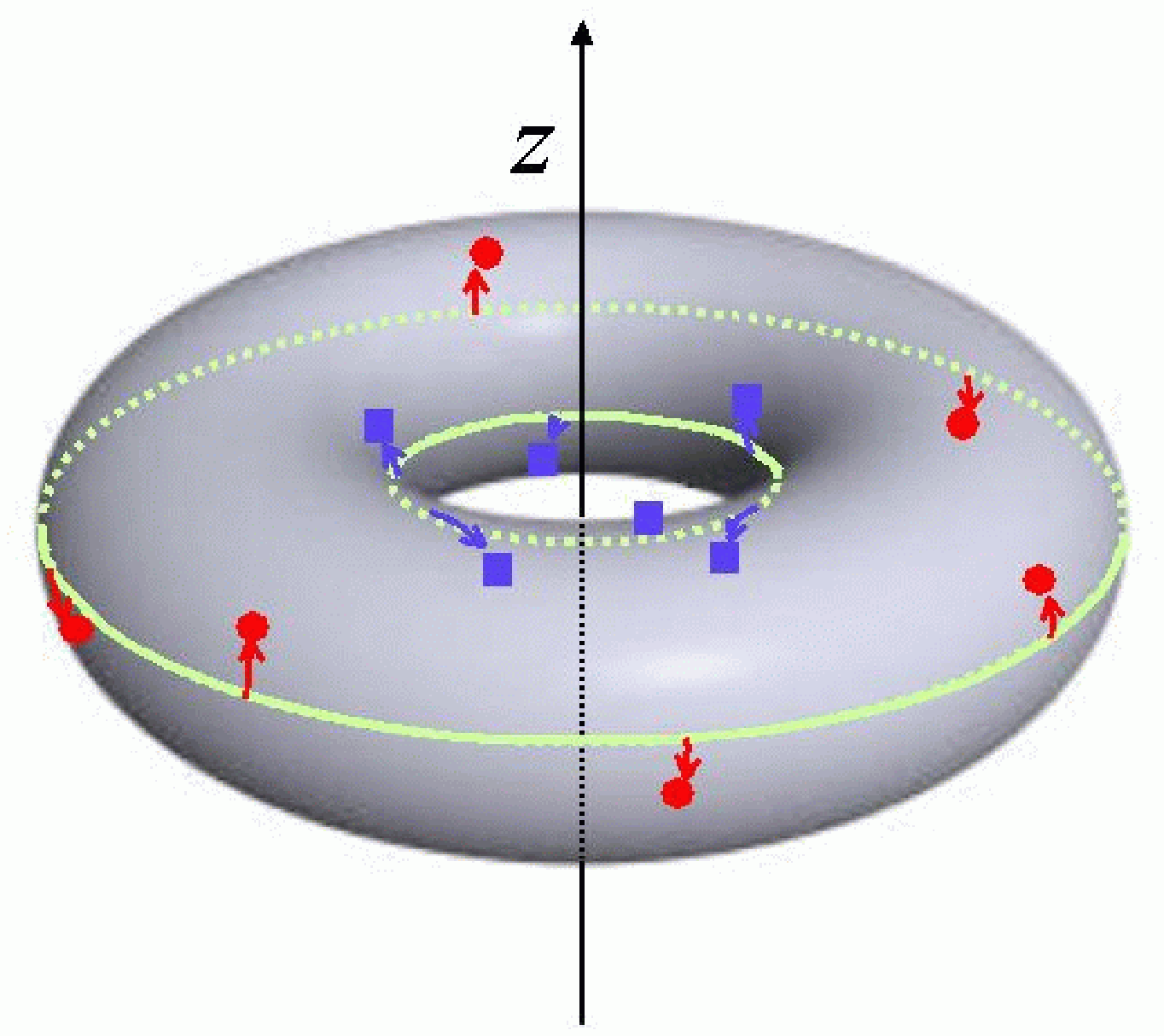}
\caption{The typical ground state configuration for parameters
that favor defect proliferation. The arrows indicate the
displacement of the equilibrium defect position from the maximal
curvature circles.} \label{torusdefects}
\end{figure}

To allow the system to reduce this energy, we have considered the
modified calculation illustrated in Fig.~\ref{fig__Ill2}. The
initial configuration starts from {\em both} circles of zero
curvature, alternating plus-minus defect pairs between them. When
the separation variable $\alpha=\pi$ the configuration is the same
as in the previous case. A typical plot of the energy is shown in
Fig.~\ref{fig__Int}. In this plot, the minimal configurations
(with $\alpha \neq \pi$) correspond to {\em buckled} or staggered
rings of defects, displaced from the circles of maximum or minimum
curvature, as illustrated in Fig.~\ref{torusdefects}. The final
position is a compromise between the ``electrostatic" repulsion
and the attraction to the Gaussian curvature basins. With the
extra degrees of freedom allowed by buckling, the energies are
always lower than the simple ring configurations of
Fig.~\ref{fig__no_core1a} and Fig.~\ref{fig__yc_core1a}. Note that
staggering allows more defects to be squeezed into the ground
state: the optimal number $N^*/2$ of defect pairs is 7 in
Fig.~\ref{fig__Int}, as opposed to $N/2 \approx 4-5$ in
Fig.~\ref{fig__no_core1a}.

\subsection{Analytical argument for defects in the ground state for aspect ratio $r$ near 1}

Let us consider a $+/-$ disclination dipole on a zero curvature
circle of the torus. Imagine slowly pulling the dipole apart until
the individual $+$($-$) disclinations reach the
outer($+$)/inner($-$) rim of the torus respectively. The total
energy in this configuration is dominated by the attraction of
each defect to the corresponding region of same sign curvature
since the defects are too far apart for the defect-defect
interaction to be important. The total energy following from
Eq.(\ref{Ising_en}) is therefore \be\label{two_L_energy} {\mathcal
E} \approx {\mathcal E}^{d-c}=-\frac{\pi}{3} K_A
\ln\left(\frac{r+1}{r-1}\right) \ , \ee where we have set
$\varphi_1=0$ and $\varphi_2=\pi$ in Eq.(\ref{bgcurvint}). Upon
approximating the defect-pair energy by its flat space value
\be\label{edd} {\mathcal E}_{dd} \approx \frac{\pi}{18}K_A
\ln\left(\frac{R_2}{a_0}\right) \ , \ee we find a total energy
\be\label{KTlike} {\mathcal E} =  -\frac{\pi}{3} K_A
\ln\left(\frac{r+1}{r-1}\right) + \frac{\pi}{18}K_A
\ln\left(\frac{R_2}{a_0}\right) + 2E_c \ . \ee If we assume,
consistent with our numerical evaluation of ${\mathcal Q}$, that
the constant correction to Eq.(\ref{edd}) is negligible, then
$E_c$ can be interpreted as a microscopic disclination core
energy.  Eq.(\ref{KTlike}) changes sign for \be\label{critsize}
\frac{R_2^c}{a_0} = \exp \left\{\frac{-36E_c}{\pi
K_A}\right\}\left(\frac{r+1}{r-1}\right)^6 \ . \ee Using
$M=\frac{8\pi^2}{\sqrt{3}}r\left(R_2/a_0\right)^2$, we conclude
that defects are favored for \be M \lesssim M_c=
\frac{8\pi^2}{\sqrt{3}} \exp \left\{\frac{-72E_c}{\pi
K_A}\right\}r\left(\frac{r+1}{r-1}\right)^{12} \ . \ee For the
representative value $E_c=0.1K_A$, we therefore find
\be\label{masterformula} M_c \approx 4.6
r\left(\frac{r+1}{r-1}\right)^{12} \ . \ee A comparison with our
numerical results for $M_c$ for both vanishing core energy and
$E_c=0.1K_A$ is shown in Fig.\ref{fig__MMax} {--} the agreement is
excellent. This result also establishes that excess defects are
present in the ground state for any fixed particle number provided
the torus is sufficiently fat. Defects could thus be an important
feature of hexatic textures for realistic vesicle sizes.

It is interesting to generalize these formulae to $p$-fold
symmetric order parameters \cite{LuPr:92} on the surface of a
torus. Here, hexatic order corresponds to $p=6$, nematic order to
$p=2$ and tilt order to $p=1$. A hypothetical ``tetradic phase"
with a four-fold liquid crystalline symmetry \cite{NH:79} would
correspond to $p=4$. The generalization of Eq.(\ref{KTlike}) for a
minimally charged defect-antidefect pair with $p$-fold symmetry
reads \be\label{KTlike_pfold} {\mathcal E} =  -\frac{2\pi}{p} K_A
\ln\left(\frac{r+1}{r-1}\right) + \frac{2\pi}{p^2}K_A
\ln\left(\frac{R_2}{a_0}\right) + 2E_c \ . \ee   This yields a
critical particle number \be M_c= \frac{8\pi^2}{\sqrt{3}}\exp
\left\{\frac{-2p^2E_c}{\pi
K_A}\right\}r\left(\frac{r+1}{r-1}\right)^{2p} \ . \ee The
critical number of particles above which defects no longer appear
in the ground state is therefore lower for coatings of the torus
by textures of lower symmetry. Since $R_1$ must exceed $R_2$ by an
amount of order $a_0$ for a physical torus (see
Fig.~\ref{fig__CO}), $(r-1)_{\rm min} = (a_0/R_2)$. Hence, $M_c$
diverges like $(R_2/a_0)^{2p}$ in the limit of an extremely fat
torus. Upon noting that $M \sim (R_2/a_0)^2$, we see that
typically $M \ll M_c$ whenever $R_2/a_0 \gg 1$. Thus, defects are
an {\em inevitable} part of the ground state for sufficiently fat
torii in all cases, except possibly for $p=1$.

\section{Temperature and shape fluctuations}\label{SECT__TEMP}

\subsection{Connection with Two Dimensional Melting}

The renormalized Frank constant for a film in the hexatic phase
has the temperature dependence \cite{NH:79}, \be\label{NHY_ka}
\frac{K_A(T)}{k_B T} \sim \frac{\xi^2_{+}(T)}{a_0^2} \ , \ee where
$\xi_{+}$ is the correlation length. The correlation length itself
behaves in the neighborhood of the hexatic to fluid transition
temperature $T_l$ like \be\label{NHY_xi} \xi_{+} \sim \exp
\left\{\frac{b}{\sqrt{|T-T_l|^{1/2}}}\right\} \ . \ee The bending
rigidity has been shown to have a much weaker temperature
dependence \cite{Klein1:86,Forster:86}. Near the hexatic-fluid
transition, therefore, the ratio of $K_A/\kappa$ diverges, which
should produce larger values of $r$.

For toroidal vesicles these results change in two important ways:
both the finite size and the Gaussian curvature of the torus must
be taken into account.  The finite area of the torus limits the
growth of the correlation length, viz. \be\label{bound_xi} \xi
\lesssim \pi R_2 \ , \ee or equivalently, \be\label{bound_KA}
\frac{K_A(T)}{k_B T} \lesssim \left(\frac{\pi R_2}{a_0}\right)^2 \
, \ee so that the Frank constant no longer diverges. It is
possible that the effects of Gaussian curvature will even limit
$K_A/\kappa$ to smaller values. As discussed in the Introduction,
it may be possible to ``freeze-in" an aspect ratio $r \approx
\sqrt{2}$ by using lipid bilayers with only short range order as a
toroidal template.

\subsection{Fluctuating Hexatic Membranes}

In \cite{DGP:87} the properties of a fluctuating hexatic membrane
were investigated. It was found that the long-distance behavior is
governed by a new fixed point, characteristic of a crinkled phase
intermediate between a crumpled and a rigid phase. Within a large
$d$ expansion, the new fixed point has the property
\be\label{DGP_rat} \frac{K_A}{\kappa}=\frac{4d}{3} \Rightarrow
\frac{K_A}{\kappa}=4 \mbox{  at $d=3$} \ . \ee It can be shown
that for the value of $r$ corresponding to this ratio of elastic
constants, additional defects should be present. The aspect ratio
as a function of the elastic constants for a defect free
configuration gives (see Eq.(\ref{sol_two})) \be\label{new_asp} r
\sim \sqrt{2} \ , \ee a Clifford torus, which we have shown
contains additional defects in the ground state.

\section{Acknowledgments}

A.T. wishes to thank S. Katz for informative discussions. D.R.N.
acknowledges conversations with V. Vitelli. The work of M.B. was
supported by the National Science Foundation under Contract
DMR-0219292 and by the U.S. Department of Energy under Contract
No. DE FG02 85ER40237. The work of DRN was supported by the
National Science Foundation under Grant No. DMR-0231631 and by the
Harvard Material Research Science and Engineering Laboratory
through Grant No. DMR-0213805. The work of A.T. was supported by
Iowa State start-up funds.

\vspace{0.25cm}

\section{Appendix}

In the angular coordinates $\{\theta,\alpha\}$ of
Fig.~\ref{fig__CO} (with $0 \leq \theta \leq 2\pi$,\,$0 \leq
\alpha \leq 2\pi$) the parametrization \bea\label{Torus_Eq1}
x&=&(R_1+R_2\cos\alpha)\cos\theta \ ,\nonumber\\
y&=&(R_1+R_2\cos\alpha)\sin\theta \ ,\\\nonumber z&=&R_2\sin\alpha
 \eea defines a torus as the locus of points $(x,y,z)$ that
satisfy $\left(\sqrt{x^2 + y^2} - R_1\right)^2 + z^2 = {R_2}^2$.
The dimensionless aspect ratio $r$ \be\label{r__def} r\equiv
\frac{R_1}{R_2} \ , \ee is constrained to be greater than one for
torii which do not self-intersect. The metric is given by
\be\label{torus__metric} ds^2=R^2_2\left\{(r+\cos\alpha)^2
d\theta^2+d\alpha^2\right\} \ . \ee Upon introducing a new angle
variable $\varphi$ ($0 \leq \varphi \leq 2\pi$) via
\be\label{new_angle} \cos \alpha =\frac{r
\cos\varphi-1}{r-\cos\varphi} \ , \ee Eq.(\ref{Torus_Eq1}) becomes
\bea\label{Torus_Eq2}
x&=&\frac{a \sinh\rho \cos\theta}{\cosh\rho-\cos\varphi} \ ,\nonumber\\
y&=&\frac{a \sinh\rho \sin\theta}{\cosh\rho-\cos\varphi} \ ,\\
z&=&\frac{a \sin\varphi}{\cosh\rho-\cos\varphi} \ , \eea where $a$
and $\rho$ are defined by $R_2=a/\sinh\rho$ and
\be\label{def_coshr} r=\cosh\rho \,
\left(\rho=\ln\left\{r+\sqrt{r^2-1}\right\}\right) \ . \ee In
these coordinates, the metric Eq.(\ref{torus__metric}) becomes
\be\label{torus_rr_mm} ds^2=R^2_2
\left(\frac{r^2-1}{r-\cos\varphi}\right)^2\left( d\theta^2+\
\frac{d\varphi^2}{r^2-1}\right) \ . \ee The metric is now
conformally flat, i.\ e.\ up to a $\varphi$-dependent
multiplicative prefactor this is the metric of a plane with
(rectangular) periodic boundary conditions.


\begin{thebibliography}{99}

\bibitem{MuBe:91}
M. Mutz and D. Bensimon, Phys. Rev. {\bf A43}, 4525 (1991); B.
Fourcade, M. Mutz and D. Bensimon, Phys. Rev. Lett. {\bf 68}, 2551
(1992).

\bibitem{MiBe:95}
X. Michalet and D. Bensimon, J. Phys. France {\bf II5}, 263
(1995).

\bibitem{SSSC:88}
G. Smith, E. Sirota, C. Safinya, and N. Clark, Phys. Rev. Lett.
{\bf 60}, 813 (1988).

\bibitem{SSSPC:90}
G.S. Smith et. al., J. Chem. Phys. {\bf 92}, 4519 (1990).

\bibitem{Sieg:99}
D. P. Siegel, Biophys. J. {\bf 76}, 291 (1999).

\bibitem{MaAl:02}
V. Markin and J. Albanes, Biophys. J. {\bf 82}, 693 (2002).

\bibitem{KoKo:02}
Y. Kozlovsky and M. Kozlov, Biophys. J. {\bf 82}, 882 (2002).

\bibitem{Helf:73}
W. Helfrich, Z. Naturforsch {\bf 28C}, 693 (1973).

\bibitem{JSL:93}
F. Julicher, U. Seifert, and R. Lipowsky, J. Phys. France {\bf
II}3, 1681 (1993).

\bibitem{Seif:97}
U. Seifert, Adv. Phys. {\bf 46}, 13 (1997).

\bibitem{LuPr:92}
T. C. Lubensky and J. Prost, J. Phys. France {\bf II2}, 371
(1992); see also R.~M.~L. Evans, J. Phys. France {\bf II5}, 507
(1995) [arXiv:cond-mat/9410010].

\bibitem{DHNMBW:2002}
A.~D. Dinsmore, M.~F. Hsu, M.~G. Nikolaides, M. Marquez, A.~R.
Bausch and D.~A. Weitz, Science {\bf 298}, 1006 (2002).

\bibitem{CMurray:1992}
C.~M. Murray, in {\em Bond-Orientational Order in Condensed Matter
Systems}, edited by K. ~Strandburg (Springer-Verlag, Berlin,
1992).

\bibitem{Scars:2003}
A.~R. Bausch, M.~J. Bowick, A. Cacciuto, A.~D. Dinsmore, M.~F.
Hsu, D.~R,. Nelson, M.~G. Nikolaides, A. Travesset and D.~A.
Weitz, Science {\bf 299}, 1716 (2003)[arXiv:cond-mat/0303289].

\bibitem{deGPro:93}
P.~G. deGennes and J. Prost, {\em The Physics of Liquid Crystals}
(Clarendon Press, Oxford, 1993). For tilt order on a torus, we
expect a maximum of two positive disclinations on the outer wall,
with compensating negative defects on the inner wall. For nematic
order, the maximum number of + or - defects is 4.

\bibitem{VN:2003}
V. Vitelli and D.~R. Nelson, to be published.

\bibitem{SeNe:89}
J. Selinger and D.~R. Nelson, Phys. Rev. {\bf A39}, 3135 (1989).

\bibitem{David:89}
F. David, in {\em Statistical Mechanics of Membranes and
Surfaces}, edited by D. Nelson, T. Piran and S. Weinberg (World
Scientific, Singapore, 1989); see also Ref.~\cite{BowTra:01a}.

\bibitem{BNT:2000}
M. Bowick, D.R. Nelson, and A. Travesset, Phys. Rev. {\bf B62},
8738 (2000) [arXiv:cond-mat/9911379].

\bibitem{BowTra:01a}
M. Bowick and A. Travesset, J. Phys. {\bf A}: Math. Gen. {\bf 34},
1525 (2001) [arXiv:cond-mat/0005356].

\bibitem{MF:1967}
See, e.g., S.B. Morse and H. Feshbach, {\em Methods of Theoretical
Physics} (Addison Wesley, New York, 1967) Vol. 1, p. 666 and Vol.
2, p.1301.

\bibitem{Polchinski:1998}
J. Polchinski, {\em String Theory}, Vol. I (Cambridge University
Press, Cambridge, 1998).

\bibitem{Mumford}
D. Mumford, {\em Tata Lectures on Theta I} (Birkha\"{u}ser,
Boston, 1983).

\bibitem{WhitWat}
E.T. Whittaker and G.N. Watson, {\em A Course of Modern Analysis},
Fourth Edition (Cambridge University Press, Cambridge, 1927).

\bibitem{NH:79}
D.R. Nelson and B.I. Halperin, Phys. Rev. {\bf B19}, 2457 (1979).

\bibitem{Young}
A. P. Young, Phys. Rev. {\bf B19}, 1855 (1979).

\bibitem{Klein1:86}
H. Kleinert, Phys. Lett. {\bf A114}, 263 (1986).

\bibitem{Forster:86}
D. Forster, Phys. Lett. {\bf A114}, 115 (1986).

\bibitem{DGP:87}
F. David, E. Guitter, and L. Peliti, J. de Physique {\bf 48}, 1085
(1987).







\end{thebibliography}
\end{document}